# A Fast Method of Virtual Stent Graft Deployment for Computer Assisted EVAR

**Aymeric Pionteck[a,b], Baptiste Pierrat[a], Sébastien Gorges[b], Jean-Noël Albertini[c] and Stéphane Avril[a]**

[a] Mines Saint-Etienne, Univ Lyon, Univ Jean Monnet, INSERM, U1059 Sainbiose, Centre CIS, F-42023 Saint-Etienne, France.

[b] THALES, Microwave & Imaging Sub-Systems, 38430 Moirans, France

[c] Univ Jean Monnet, INSERM, U1059 Sainbiose and University Hospital of Saint-Etienne, F-42000 Saint-Etienne, France

**Abstract** In this paper we introduce a new method simulating stent graft deployment for assisting endovascular repair of abdominal aortic aneurysms. The method relies on intraoperative images coupled with mechanical models. A multi-step algorithm has been developed to increase the reliability of simulations. The first step predicts the position of the stent graft within the aorta. The second step is an axisymmetric geometric reconstruction of each individual stent. The third step minimizes the rotation of each stent around its main axis. Finally, the last step virtually deploys each stent within a deployment box extracted from the preoperative CT scan. A proof of concept is performed on a patient. The accuracy is compatible with the clinical threshold of 3 mm: the average distance between target and simulated stents is $1.73 \pm 0.37$ mm. Fenestrations of the stent-graft are reconstructed with a maximum error of less than 2.5 mm, which is enables a secure catheterization of secondary arteries. In summary, the method is able to assist EVAR practitioners by providing all necessary information for a fast and accurate stent graft positioning, combining intraoperative data and a mechanical model in a very low cost framework.

# I. INTRODUCTION

Abdominal aortic aneurysm (AAA) is a frequent asymptomatic pathology that results in abnormal local deformation of the aorta. Each year, aneurysm ruptures are responsible for 10,000 deaths in the United States [1]. Clinical monitoring of the evolution of the aneurysmal sac diameter is used to decide whether an intervention is necessary [1], [2]. Two options are available: conventional open surgery or endovascular surgery (EVAR). Endovascular surgery is associated with a lower mortality rate (1.5%) than open surgery (4.6%), although long-term mortality is similar [3], [4].

During EVAR, the surgeon first makes a small incision in the groin to reach the femoral artery. From this incision, tools are introduced to position the stent graft (SG) launcher within the aneurysm. Then the SG is progressively deployed. The success of the intervention depends on the precise positioning of the SG in the artery. In some cases, a fenestrated SG is required if the aneurysm extends beyond the ostia of the renal arteries. In this case, the fenestrations of the SG must be positioned precisely in front of the renal ostium whose diameter is about 5-7mm. This phase is delicate but essential to allow the catheterization of the secondary arteries and avoid occlusions and post-operative complications [5], [6] [7], [8]. The lack of 3D information obliges the surgeon to perform a mental reconstruction of the scene using several images with different incidence angles, which considerably increases the duration of the procedure, as well as the time of exposure to X-rays and the volume of injected contrast products. Therefore, the virtual 3D representation of the tool location and particularly the SG in the aorta is a valuable aid to the surgeon. This would reduce the surgery time and the number of X-ray images required. Moreover, it would reduce the number of postoperative complications, most often related to inaccurate SG positioning. Latest generation systems enable the acquisition of 3D images of the tool during the intervention [9], [10]. Recently, efforts have been made to use biplanar fluoroscopic acquisitions to reconstruct the 3D shape of the device [11]–[14]. However, all these methods are based on expensive equipment which are not commonplace in all hospitals, usually equipped with simple mobile C-arms. Another solution should therefore be available to obtain a three-dimensional representation of the inserted SG at low cost. Modelling and numerical simulation of SG deployment then appears as essential.

First studies on numerical simulation of SG deployment were based on finite element analyses to study the mechanics of stents and to simulate their deployment in arteries [15]–[17], integrating different types of constitutive behavior for the different materials of SGs [18], [19], [20].

Perrin et al. [21]–[24] developed a preoperative planning tool to predict the postoperative position of the SG from patient-specific models. Although essential for preoperative planning, these studies have two important limitations with regard to their use as real time assistance for the practitioner: (i) inappropriately long computation time, and (ii) lack of update from intraoperative images. Some studies have focused on reducing the computation time and developed algorithms to simulate stent deployment in "real time". They often rely on simplifications such as modeling vessels as generic tubes, on which the stent armatures are then mapped [25]–[27]. The Fast Virtual Stenting (FVS) technique was proposed by Larrabide et al [28]. This technique is based on constrained deformable simple models and can virtually model stent deployment in vessel and aneurysm models. The FVS technique was tested and compared with experimental results [29] and with finite element models [30][31]. Alternative methods have been proposed, based on mass-spring models [32], [33] or on active contours [34]. Although efficient and fast, most of these models are based on simplified mechanics and can be challenged by complex vascular geometries. In addition, applications focus on preoperative planning, as none of the work mentioned above considered intraoperative images.

A small number of studies integrated information from intraoperative images. For example, Demirci et al.[35] proposed an algorithm to automatically match a 3D model of the SG with an intraoperative 2D image of its structure. Zhou et al.[36] , and Zheng et al. [37]  introduced a real-time framework to generate the 3D shape of a fenestrated SG from a single 2D fluoroscopic image and position of added radio-opaque markers. These methods have reduced computation times and can accurately represent the deployment of SGs in simple geometries. However, more complex cases cannot be addressed without the use of a mechanical model.

To our best knowledge, no studies have ever combined these different aspects into a single method. Achieving this combination is the objective of the present work, in order to propose a method that can assist EVAR practitioners by providing all necessary information for a fast and accurate SG positioning.

The details of the method are given in this book chapter, first introducing the global algorithm, then describing each step and finally showing a proof of concept for a patient case.

## II.   METHODS

The global algorithm of the method is summarized in Figure 1. The input data are the 2D intraoperative images from a mobile C-arm and the 3D geometry of the aorta obtained from a preoperative CT scan. The algorithm is divided into four main steps. The first two steps can be combined into a single stage called Stage 1. This stage is essential for the following steps but may reach insufficient accuracy, hence the possible following Stage 2. During the first step of Stage1, barycenters of each stent are positioned in 3D using a FEM model of the SG in the aorta. Then, the stents are geometrically reconstructed during the second step of Stage 1. If necessary, two refining steps  are achieved during Stage 2, which is an updating or refining stage. These supplemental steps require a slightly longer calculation time but reach higher accuracy. The first step of Stage 2 consists in recovering the rotation of the stent around its main axis through a minimization loop. The second step of Stage 2 consists in deploying each stent individually.

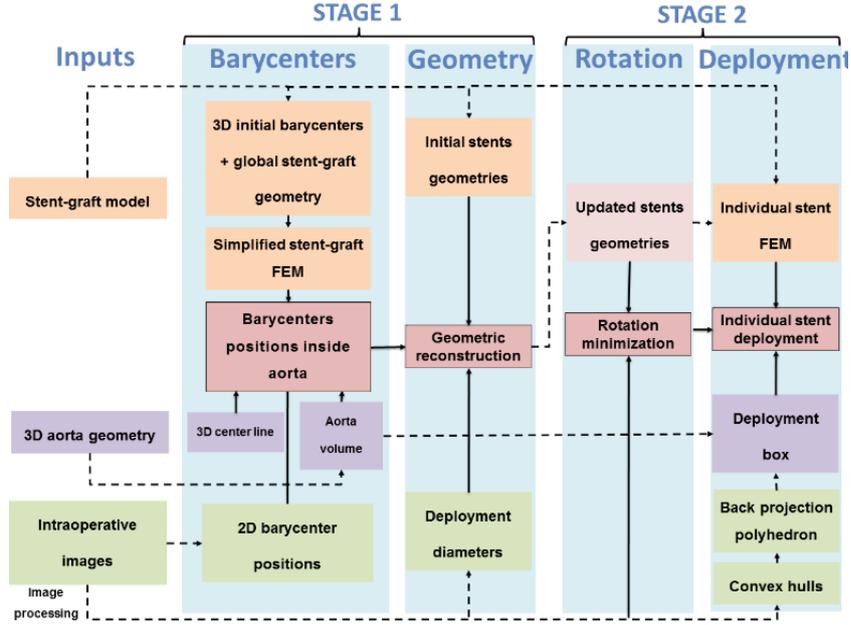

**Figure 1 – Schematic of the general algorithm**

## A. Data acquisition

In this section, we list the input data and describe how they are processed to extract relevant information and feed the simulations. Data available at the beginning of SG reconstruction include the SG model, updated 3D aorta geometry and intraoperative imaging. The SG models are obtained using the method described in [22], [38]. Briefly, stent geometries are obtained from manufacturer specifications and are discretized into finite elements using a dedicated Matlab® routine. All simulations are constrained and guided by intraoperative imaging. To isolate the contour of the stents, a combination of Frangi filters [39] and masks is applied to the image [35]. The Frangi filter is generally used to detect vessels or tubular structures in volumetric image data. The Frangi filter is available in open-source libraries and software such as ITK or ImageJ. The filter includes a measuring scale that allows the isolation of tubular structures of different sizes. By modifying the scale of the filter and combining it with masks, it is possible to extract binary tubular structures of the stents (Figure 2). Then, the convex hull of each stent is extracted. The two-dimensional coordinates of stent barycenters are simply obtained from the convex hull (Figure 2). Apparent deployment diameters are also measured. For each stent, the proximal diameter $d_p$ and distal diameter $d_d$ are recorded. They will be used for the further geometric reconstruction of stents.

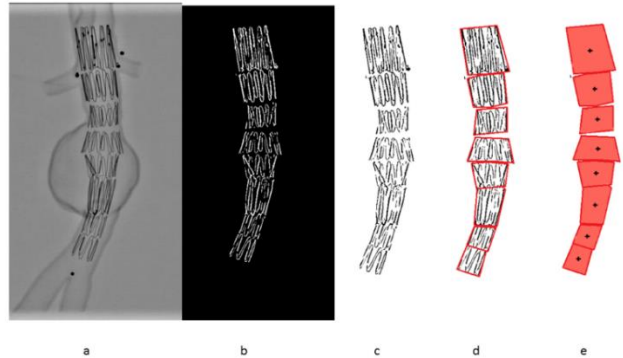

**Figure 2 – Example of stent deployment in a 3D printed AAA replica: original image (a), stent detection (b) and extraction (c), convex hull (d) and 2D barycenters (e)**

In the following steps, the SG is virtually positioned in the 3D geometry of the aorta. The aortic geometry, including the centerline and the volume, is previously extracted from the preoperative scan. The volume of the aorta is segmented from the DICOM file of the preoperative scan with a front collision method implemented in VMTK [40]. Then, the centerline is extracted using the Voronoi diagram method, also implemented in VMTK. The geometry of the aorta obtained from the preoperative scans may be slightly different of the aortic geometry at the day of the intervention. Indeed, it can be deformed, especially when stiff guidewires are inserted. The aortic geometry must be updated before simulating SG deployment. To do so, the geometry is rigidly and then non-rigidly registered on the intraoperative images. Several methods are available for this step [41] [42] ([43] under review). In addition, we assume that we know the projection matrix of the C-arm.

## B. Corotational Euler-Bernoulli beam elements

In this section we describe the corotational Euler-Bernoulli beam elements that are used in the following steps to discretize the simplified geometry of the SG in the global positioning step, and then the stents in the individual deployment step. Simulations are carried out with Project Chrono [44], [45]. The details of the theory and implementation of beam elements are described in [46]. We review here the main concepts. Among the different methods that allow simulating large deformations by finite elements, the corotational approach is one of the most versatile as it is based on classical linear finite elements. The corotational approach allows large displacements, but requires that the strains remain small (Figure 3). A floating coordinate system $\mathbf{F}$ follows the deformed element, so that the overall movement in the deformed $\mathbf{C_D}$ state can be assumed to be composed of a large rigid body movement from the reference configuration $\mathbf{C_0}$ to the so-called floating or phantom configuration $\mathbf{C_S}$, times a small local deformation from $\mathbf{C_S}$ to $\mathbf{C_D}$. The underlined symbols represent variables expressed in the floating reference basis $\mathbf{F}$. A global tangent stiffness $\boldsymbol{K_e}$ and a global force vector $\mathbf{f_e}$ are derived for each element $\mathbf{e}$, given its local matrix $\underline{\boldsymbol{K}}$, its local force $\underline{\mathbf{f}}$ and the rigid body motion of $\mathbf{F}$ in $\mathbf{C_0}$ to $\mathbf{F}$ in $\mathbf{C_S}$. At each time step, the position and rotation of $\mathbf{F}$ are updated.

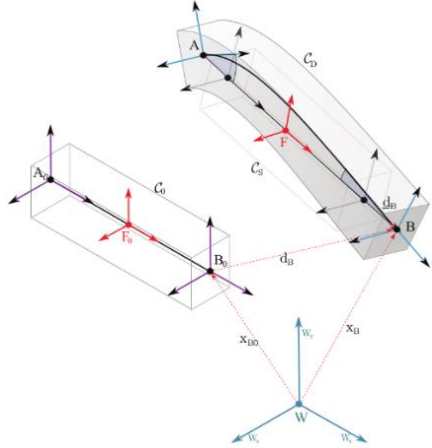

**Figure 3 - Schematic representation of the corotational approach**

## C. Stage 1 (preliminary stage)

Stage 1 combines the first two steps of the algorithm: positioning of the stent barycenters in the aorta and axisymmetric reconstruction of the stents. This step can be run in real-time as it has a marginal computational cost. However, it is based on assumptions that may not be fully satisfied in practice. Thus, the output of this preliminary stage will serve as the starting-point for the updating stage presented in the next section. The first step is to recover the global position of the SG inside the aorta. The SG is simulated with a simplified finite element model. The algorithm for positioning barycenters in 3D is summarized in Figure 4.

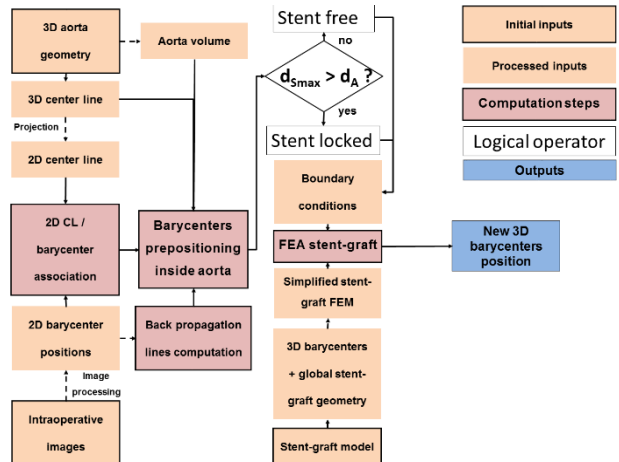

**Figure 4 - Overview of the algorithm for 3D positioning of barycenters**

## Barycenter positioning

It is very challenging to find the position of the SG directly in the global reference frame from a single image. Indeed, the intrinsic nature of the C-arm conical projection and the discretization in pixels of the detectors lead to a significant incertitude along the projection axis. A pixel from the flat panel can be assimilated to a surface, therefore its back projection geometry is not a line but a pyramidal volume. It is from this volume that the uncertainty on point positioning comes from (Figure 5). This uncertainty represents the distance along which an object can be moved along the projection axis without moving in the projection plane. This uncertainty depends on the position of the object, i.e. the source-object distance but also the distance from the projection axis. When the pixel gets closer to the projection axis, the uncertainty may tend towards infinity. Close to the edges of the image, the uncertainty becomes lower. For a standard case (source-object distance = 800 mm, source-detector distance = 1300 mm, 750 pixel x 750 pixel detector), a point located 175 pixels from the projection axis has an uncertainty of 4.6 mm.

However, our objective is to reduce this uncertainty as much as possible. Accordingly, our method uses the aortic geometry as a support for the overall SG positioning. The first step is to relate each stent barycenter with the closest point of the artery centerline. This association is achieved in two dimensions. The centerline of the aorta is projected according to the same projection parameters as the intraoperative image. Each two-dimensional barycenter is simply associated with the nearest projected centerline point. This associativity is converted into three dimensions, assuming that the nearest 2D point is almost equivalent to the nearest 3D point. Each barycenter is then related to the corresponding centerline position **CL** $(x_{CL}, y_{CL}, z_{CL})$.

The next step is to obtain the back-projection lines of each 2D barycenter. For each stent, we know (from previous image processing) the convex hull and the 2D position of its barycenter $\mathbf{B_{Im}}$ $(U_{Im}, V_{Im})$ in the screen frame. The position of the target image in the 3D space, representing the configuration of the X-ray source and the flat panel detector, is known. Thus, each barycenter on the image can be associated with a three-dimensional position $\mathbf{B_{2D}}$ $(x_{2D}, y_{2D}, z_{2D})$. Since we know the projection matrix, we can obtain the corresponding back projection line for each barycenter. These lines are combined with the geometry of the aorta to obtain the global position of the SG in the 3D frame. Equations of the projection lines result from the elementary Cartesian geometry. The X-ray source is at the origin of the global coordinate system. Therefore, all projection lines have point **O** **(0,0,0)** in common. The projection parameters are known. Thus, each $\mathbf{B_{2D}}$ target point of the image is associated with a back-projection line passing through this point and through the origin (Figure 5). The normalized vector of line $\vec{v}(v_x, v_y, v_z)$ is:

$$\vec{v} = \frac{\overline{B_{2D}O}}{\|\overline{B_{2D}O}\|} \tag{1}$$

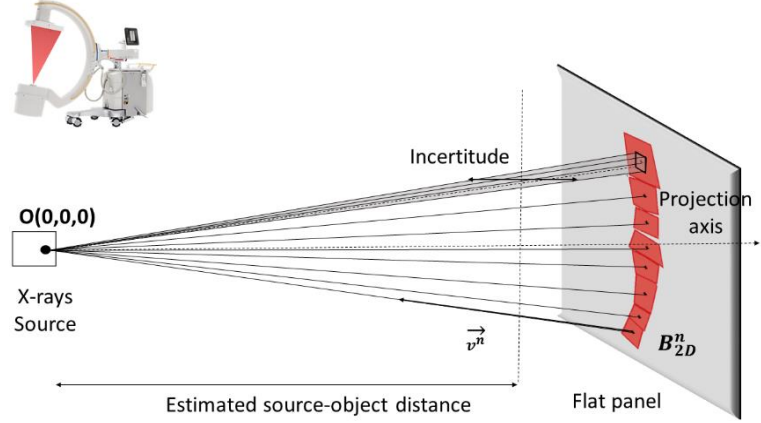

**Figure 5 - Back projection lines of 2D barycenter coordinates, according to the configuration of the mobile C-arm and uncertainty of positions along the projection axis at an estimated source-object distance**

The SG must now be pre-positioned inside the aorta. The coordinates of each three-dimensional barycenter $\mathbf{B_{3D}}$ ($\mathbf{x_{3D}}$, $\mathbf{y_{3D}}$, $\mathbf{z_{3D}}$) have then a back-projection line and have previously been associated with the closest centerline point $\mathbf{CL}$ ($\mathbf{x_{CL}}$, $\mathbf{y_{CL}}$, $\mathbf{z_{CL}}$). We know that each barycenter 3D coordinate is located on its back projection line, but the coordinate $p$ (Equation 2) of $\mathbf{B_{3D}}$ along its line is **initially** unknown.

$$\mathbf{B_{3D}(x_{3D}, y_{3D}, z_{3D}) = B_{2D}(x_{2D}, y_{2D}, z_{2D})} + p * \vec{\mathbf{v}}(\mathbf{v_x}, \mathbf{v_y}, \mathbf{v_z}) \qquad (2)$$

As a first approximation, we assign to each barycenter the coordinate $p$ from the nearest centerline point such as $\mathbf{z_{3D}}=\mathbf{z_{CL}}$. From equation 2, we obtain the following system of equations:

$$\begin{cases} \mathbf{x_{3D}} = \mathbf{x_{2D}} + p * \mathbf{v_x} \\ \mathbf{y_{3D}} = \mathbf{y_{2D}} + p * \mathbf{v_y} \\ \mathbf{z_{3D}} = \mathbf{z_{2D}} + p * \mathbf{v_z} \end{cases} \qquad (3)$$

Hence:

$$p = \frac{\mathbf{z_{3D}} - \mathbf{z_{2D}}}{\mathbf{v_z}} \qquad (4)$$

($\mathbf{x_{3D}}$, $\mathbf{y_{3D}}$) are calculated by solving the system of equations (3). Then, we have a first approximation of the position of barycenters, based on the centerline. In some cases, this approach is insufficient and must be completed using a finite element model of the SG. Figure 6 shows what can happen in the case of a large aneurysm sac. In this case, the centerline follows the shape of the artery. With the approach described above, stents would be positioned in the sac, which is unlikely and mechanically unrealistic.

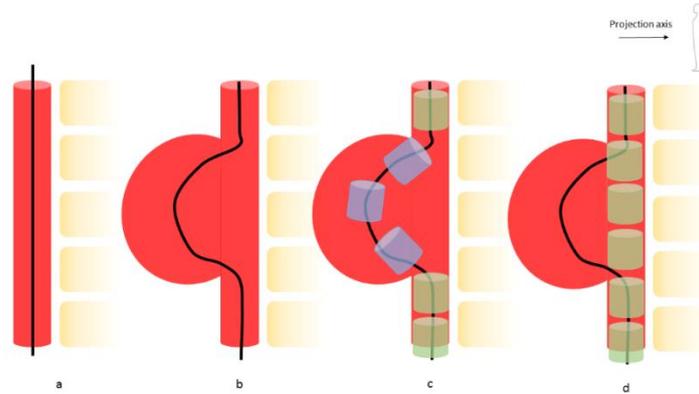

**Figure 6 – Error in barycenter positioning due to a deformed centerline: normal centerline (a), centerline deformed in an aneurysm sac (b), resulting unrealistic stent positioning (blue) in the sac area and realistic positioning (green) in non-deformed sections (c), expected actual positioning (d)**

Therefore, stents that are likely to be badly positioned must be separated from the other ones. In order to define the most precise boundary conditions for the finite element simulation, stents are divided into two categories: free and locked stents. The maximum diameter $d_{Smax}$ of the stent, i.e. the diameter of the fully deployed stent, is compared to the local diameter of the aorta $d_A$ at the associated point of the centerline. If $d_{Smax} > d_A$, the stent is locked. In this case, we assume that the stent is in equilibrium in the artery and that its barycenter is therefore very close to the local center of the artery, and therefore to **CL**. If $d_{Smax} < d_A$, for example if the stent is in the aneurysm sac, the stent is free. The position of the locked stents is assigned according to the centerline (see previous section). The position of the barycenters of free stents will be calculated using a simplified finite element model of the SG. From the initial three-dimensional geometry of the stent, the 3D position of each stent barycentre is extracted. Barycenters are connected between each other by co-rotational Euler-Bernoulli beam elements according to the initial configuration of the SG [46] (Figure 7). The model is set up using the Project Chrono libraries.

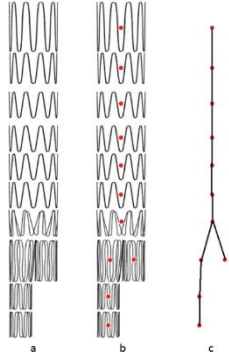

**Figure 7 - From the simplified geometry of the SG (a), 3D coordinates of barycenters are first extracted (b), then barycenters are connected using beam elements (c)**

The SG model in its initial configuration is pre-positioned in the aorta. Then, displacements are prescribed onto the locked stents and the resulting displacements of the free stents are calculated. Free stents cannot go outside the aortic lumen. A SG is a tubular structure with a high degree of mechanical inhomogeneity due to the combination of metal stents and textile graft. The SG model is very simplified, reducing the SG model to a succession of beam elements with the same mechanical behavior. The mechanical characteristics of these beams therefore have no physical reality, and have been optimized to ensure the robustness and stability of the model. As the model is subject to successive boundary conditions (back-projection lines, aortic volume), we assume that this simplified model is sufficient for our application, while allowing a very short computation time. The 3D positions of barycenters are finally determined, hence stent orientation.

### Geometric stent reconstruction

From the updated 3D position of the barycenters, a geometric reconstruction of the stent is performed. The initial geometry of each stent, i.e. the metal structure, is first discretized into a set of points (Figure 8). Each point is defined as a vector $\underline{\mathbf{V}}(\mathbf{V_x}, \mathbf{V_y}, \mathbf{V_z})$, which originates from the stent barycenter $\mathbf{B_{3D}}$ and is expressed in the stent local coordinate system. Initially, the local reference coordinate system is the translated global coordinate system. Therefore, the coordinates $\mathbf{S(x,y,z)}$ of the $n$ points of a stent are defined by :

$$\mathbf{S}^{i:1 \rightarrow n} = \mathbf{B_{3D}^i} + \underline{\mathbf{V}^i}$$ (5)

The proximal deployment diameters $d_p$ and distal $d_d$ are measured during the image processing step. Here we assume that the stent deployment is axisymmetric. Thus, the diameter measured in the plane of the image is assumed to be the same in all directions. The local deployment diameter $d$ of the stent is therefore interpolated along its main axis $\mathbf{z'}$, initially coinciding with the axis $\mathbf{z}$ of the global reference frame (Figure 8). New reconstruction vectors $\overrightarrow{\mathbf{V_d}}(\mathbf{V_{d_x}}, \mathbf{V_{d_y}}, \mathbf{V_{d_z}})$ are updated according to the diameter reduction $rd$ such as:

$$rd = 1 - \frac{dSmax - d}{dSmax} \tag{6}$$

$$\begin{cases} \mathbf{V_{d_x}} = \mathbf{V_x} * rd \\ \mathbf{V_{d_y}} = \mathbf{V_y} * rd \\ \mathbf{V_{d_z}} = \mathbf{V_z} \end{cases} \tag{7}$$

Reconstruction vectors $\overrightarrow{\mathbf{V_d}}$ in the global reference frame is expressed in the global frame with the rotation matrix $\boldsymbol{R}$ according to:

$$\begin{pmatrix} \overline{\mathbf{V_d}} \\ 1 \end{pmatrix} = \boldsymbol{R} \begin{pmatrix} \overline{\mathbf{V_d}} \\ 1 \end{pmatrix} \tag{8}$$

Finally, the new position of the stent is calculated with equation 5 from the updated vectors $\overrightarrow{\mathbf{V_d}}$. The SG is eventually reconstructed (Figure 8).

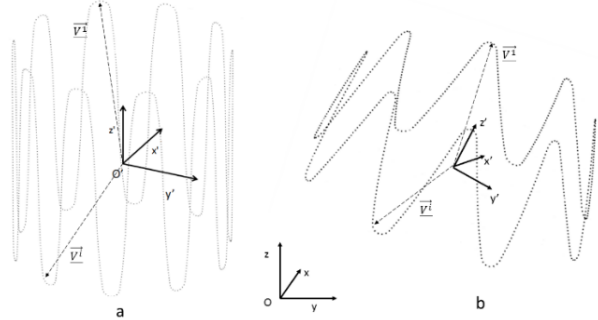

**Figure 8 – Stent geometric reconstruction, with the local frame O'(x',y',z')
and the global frame O(x,y,z), initial model (a) and after reconstruction (b)**

Positioning and reconstruction of stents is based on two assumptions: the center of gravity of the stents is close to the centerline of the aorta and the deployment of the stents is axisymmetric. Results of Stage 1 (III.C) show that these assumptions are a source of uncertainty during stent reconstruction, which may prevent in some cases to correctly simulate stent deployment. Additional steps of individual stent modeling and deployment corrections are therefore required.

### D. Stage 2 (refining stage)

Individual stent deployment may not be accurate enough and need to be improved. This stage combines two individual refining steps: minimization of rotation and individual deployment of stents. The goal of the first step is to determine the actual angle of rotation $\Phi$ of each stent, around its main axis $\mathbf{z'}$. The aim of the second step is to individually simulate the deployment of the stent in a deployment box. Both have clinical applications. When the SG is deployed, the actual SG deployment can be reconstructed in 3D. When the SG is not fully deployed, the EVAR practitioner can visualize how the SG would deploy at its current position.

### Minimization of the rotational difference

During deployment, the stent may be subject to rotations $\Phi$ around its main axis $\mathbf{z'}$. In the case of axisymmetric stents, this rotation has little impact on its final deployment, although the surgeon may wish to improve the accuracy of reconstruction for critical stents. However, in the case of stents with fenestrations, their positioning depends on the rotation of the stent. It is therefore essential to determine these rotations. This step is performed within a minimization loop with a differential evolution algorithm.

In the case of axisymmetric stents, the value to be minimized is the difference $e_S$. This difference is calculated in two dimensions. The 3D model of the stent geometrically reconstructed at the end of the previous step is projected according to the projection parameters of the target image. $e_S$ is the average distance between each point of the reconstructed stent and its nearest neighbor. This loop is used to determine the proper rotation $\Phi \pm k\theta$, where $\theta$ is the periodic angle separating two peaks of the Z-shape axisymmetric stent and $k$ a real integer. Considering fenestrated SG, all stents with fenestrations or scallops have radiopaque markers to guide the positioning. The new $e_M$ deviation to be minimized is calculated by considering only the distance between the radiopaque markers. In this case, the proper rotation $\Phi$ is exact and does not depend on $\theta$. Indeed, the positioning of fenestrations is asymmetrical.

### Individual stent deployment

The objective of this deployment box is to make maximum use of intraoperative image data. Thus, the deployment of the stent will be constrained not only by the geometry of the aorta, but also by the information from its convex hull on the image. From the convex hull obtained previously, we define a back-projection polyhedron (BpP). Each side of the BpP is a triangular element. The X-ray source is at the origin of the global coordinate system. Therefore, all triangular elements have a common point $\mathbf{O}$ $\mathbf{(0, 0, 0)}$. Each edge of the convex hull is defined as the edge opposite to the apex $\mathbf{O}$ (Figure 9). The volume is closed by the surface of the convex hull discretized in triangular elements. The Boolean intersection of the BpP with the aorta gives a volume called the deployment box. Boundaries of the volume are meshed with rigid shell elements. The stent is positioned in this rigid box according to its previously determined configuration.

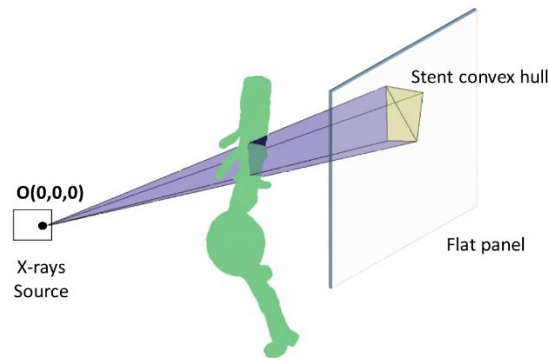

**Figure 9 – Deployment box extracted from the intersection of the aorta volume and the BpP**

The stents are composed of corotational Euler-Bernoulli beam elements. The mesh size is refined in high curvature areas. The stents are modelled in their 3D position and orientation determined in the previous steps. Only the diameter of the stent is changed. The stent model is initialized in its deployed configuration. Then it is pre-constrained to the diameter of the SG launcher before the simulation begins. Euler-Bernoulli beam elements have an elastic linear behaviour. The stents have a diameter of 0.125mm and are made of 316L steel which mechanical characteristics are summarized in [38].

Each stent, or a selection of stents at the practitioner's discretion, is deployed individually. The crimped stent is positioned inside the deployment box. The position of its barycenter and its orientation have already been recovered during the global positioning step. Then the stent is deployed (elastic recoil as the stent was crimped). The deployment is calculated using the Project Chrono engine [45], with solver Math Kernel Library (MKL) from Intel®. After the first contact, the time step is reduced to ensure the stability of the model. The contacts are modelled using the penalty algorithm implemented in Project Chrono, the Smooth-Contact (SMC) modeling approach. SMC uses penalty (in a discrete element method (DEM) [47], [48], regularizing the frictional contact forces, with "imaginary" spring-dashpot systems at each contact) and as such objects in contact will have slight interpenetration and integration time-step will likely be small. The simulations are performed on a computer with 4 CPUs, 3.40GHz, 16 GB RAM, but without parallelization. The computation time for deploying a stent is less than 6 minutes, without optimization. In addition, the complete calculation is easily parallelizable, as each stent deployment can be simulated on a separate core. The overall calculation time could therefore be compatible with clinical use.

# III.   PROOF OF CONCEPT

The method described above was applied to a patient who underwent an EVAR procedure.

## A.   Clinical data

Details about the device are given in Tab. 1. An additional stent was tested, including three fenestrations: right renal artery, mesenteric artery and left renal artery fenestrations. The preoperative and postoperative CT scans were acquired at the Saint-Etienne University Hospital under clinical conditions. Use of the clinical data was approved by the institutional review board and informed consent was obtained from the patient. The voxel size of the scans was 0.9395x0.9395x2 mm³.

**Table 1 - Clinical summary of EVAR procedure**

| | |
|---|---|
| Sex | Male |
| Age (y) | 78 |
| Stent-graft : | ENBF-28-20-C-170-EE |
| - Number of stents | 20 |
| - Proximal Graft Diameter (mm) | 28 |
| - Distal Graft Diameter (mm) | 20 |
| Anevrismal sac thrombus: | No |

First, we evaluated the results of Stage 1. As the results were not precise enough in terms of radial expansion, a second step of individual stent deployment was required. We evaluated the results of the complete method, including Correction Part. The average diameter of the renal arteries was 5 to 7 mm. If the fenestration positioning error is less than 3 mm, the surgeon can catheterize the secondary arteries such as renal arteries. Above this threshold, it is considered that intraoperative complications are likely to arise. Therefore, the clinical validation value was set at 3mm in accordance with experienced clinicians.

## B.   Quality assessment of stent deployment

First, to isolate and test the stent reconstruction algorithm as precisely as possible, the following assumptions were made: the projection matrix was known; the 3D geometry of the aorta was assumed to be perfectly registered. This ensured that not introducing positioning errors related to aortic registration. Thus, target images and the 3D model of the aorta were generated from the postoperative scan. The actual 3D position of the SG was therefore known, and served as a reference to be compared with the simulation for the sake of validation (Figure 10).

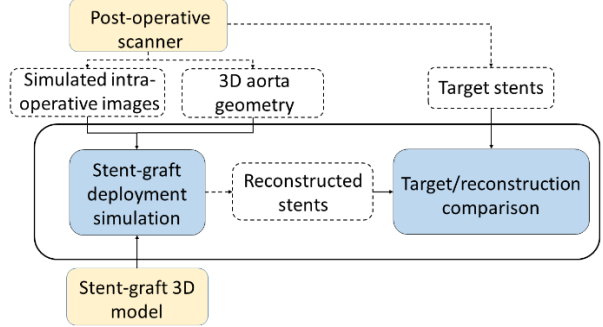

**Figure 10 – Flowchart of the validation scheme**

Several parameters were used to assess the quality of the reconstruction. The first parameter was the $D_B$ distance, which was the distance between the 3D barycentres of the target stent and the reconstructed stent. This distance was used to assess the quality of stent positioning within the artery. The second parameter was the distance $D_M$, which was the average distance between the point clouds of the target stent and the reconstructed stent. It was defined as the average of the Euclidean distances between a node of the reconstructed stent and its nearest neighbor among the points of the target stent. This distance allowed reaching the quality of stent positioning and deployment to be assessed at the same time. Finally, the last parameter is $CSAS$, i.e. the cross-sectional area overlap of target and reconstructed stents, which allows the quantitative comparison of stent deformations at the SGs folds. Cross-section areas along the entire stent were measured after deployment in the aneurysm. After deployment, the cross-sectional areas were modified, particularly in terms of SG folds. Let $A_T$ be the area of the target cross-section $S_T$ and $A_R$ the area of the reconstructed cross-section $S_R$. $A_U$ is the area of the $S_U$ intersection between $S_T$ and $S_R$.

$$S_U = S_T \cup S_R \qquad (9)$$

$$CSAS = 100 * \frac{A_T - |A_T - A_U|}{A_T} \ (\%) \qquad (10)$$

## C.   Results and discussion

Figure 11 shows the results of Stage 1 and Stage 2 and Table 2 shows a comparison of the results of the two stages. Concerning Stage 1, the average $D_B$ is generally lower than the clinical validation value, the positioning of stents in the artery is generally good. However, the error is too large on the contralateral limb, close to or greater than 3 mm with a maximum of 3.71 mm. About $D_M$, the average error is less than 3 mm, however the distance map shows that a significant number of stents have an error too close to the limit. The $CSAS$ map confirms this result, with an average superposition of about 70% and a minimum of 47%, which is much too low. Thus, an additional step of individual deployment seems necessary. Stage 2 showed a clear improvement in the quality of stent simulation. All measurements were below the threshold value, with a maximum for the $D_B$ of 1.40 mm and for the $D_M$ of 2.28 mm. The mean values were also improved, by reducing the mean error of -26.5% for the $D_B$ and -21.7% for the $D_M$ compared to Stage 1. There was also a clear improvement in the *CSAS* by a mean value of 25%.

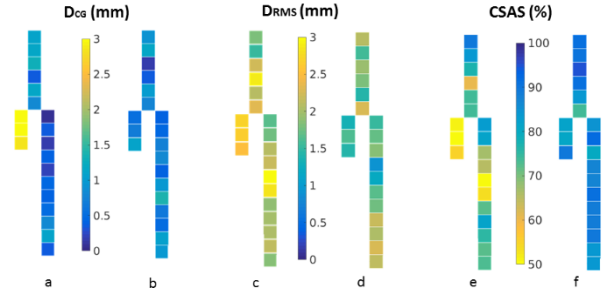

**Figure 11 – Distance map representing the distance between the barycenters of the target and the reconstructed SG from Stage 1 (a) and Stage 2 (b), the RMS distance between the cloud points of the target and the reconstructed stents from Stage 1 (c) and Stage 2 (d), and the cross-sectional surface superposition between target and reconstructed stents from Stage 1 (e) and Stage 2 (f). Each square represents a stent.**

**Table 2 - Summary and comparison of Stage 1 and Stage 2 performances**

|  |  | Preliminary Part | Correction Part | Difference | Difference (%) |
|---|---|---|---|---|---|
| $D_B$ (mm) | Mean ± std | 1.02 ± 1.03 | 0.75 ± 0.32 | -0.27 | -26.5 |
|  | Max | 3.71 | 1.40 | -2.31 | -62.3 |
|  | Min | 0.05 | 0.15 | 0,1 | 200.0 |
| $D_M$ (mm) | Mean ± std | 2.21 ± 0.44 | 1.73 ± 0.37 | -0.48 | -21.7 |
|  | Max | 3.12 | 2.28 | -0.84 | -26.9 |
|  | Min | 1.54 | 0.90 | -0.64 | -41.6 |
| *CSAS* (%) | Mean ± std | 69.2 ± 12.8 | 86.6 ± 5.1 | 17.4 | 25.1 |
|  | Max | 89.3 | 95.2 | 5.9 | 6.6 |
|  | Min | 46.9 | 73.7 | 26.8 | 57.1 |

Figure 12 shows how fenestration positions (front view) were predicted. The size difference between the target and reconstructed fenestrations results from the segmentation of the post-operative SG. The fenestrations are surrounded by radio-opaque markers that create artifacts during imaging. These artifacts make the markers appear larger than reality. The average distances between the centers of gravity of the fenestrations are summarized in Table 3. The overlapping of fenestrations is improved after the refining steps, which is visible on the Figure 12. The method, after correction, was therefore able to position the fenestrations precisely enough to allow the catheterization of secondary arteries, with a maximum distance less than 2.46 mm for the left renal fenestration.

These results are interesting in comparison with previous work. Indeed, preoperative finite element simulations [21] have an average error of about 3.5mm, which is higher than the results obtained here, with a longer computation time, as they are intended for planning purposes. It is though difficult to compare the previous fast methods with the one presented here. Indeed, the evaluation criteria are not similar. For example, in [33], the results of the fast method (FM) were compared with finite element simulations (FE), which may themselves differ from reality. In complex cases, the average distance between FM and FE is about 3-4 mm, for a very short calculation time (<1min). However, these methods are not guided by intraoperative imaging. In [36], an accuracy of about 2-3 mm is reached, which is comparable to our results, with a shorter computation time. However, their method was based on additional markers bonded onto the device, which does not seem suitable. Compared to these previous studies, our method seems to achieve a promising compromise between accuracy, computation time and compatibility with current clinical conditions.

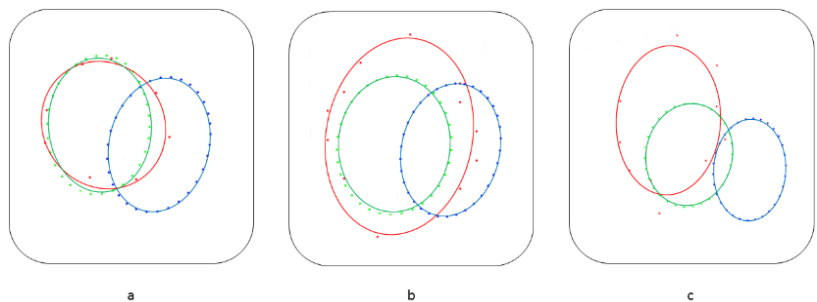

**Figure 12 - Comparison of the positions of the fenestrations. In red, the target, in blue the reconstructed fenestration after Stage 1, in green after Stage 2. Right renal fenestration (a), mesenteric fenestration (b) and left renal fenestration (c)**

**Table 3 - Distance between the center of gravity of target fenestrations and reconstructed fenestrations, before and after Stage 2**

|  | Right renal fenestration | Mesenteric fenestration | Left renal fenestration |
|---|---|---|---|
| $D_B$ Stage 1 (mm) | 3.06 | 3.05 | 6.05 |
| $D_M$ Stage 2 (mm) | 0.41 | 1.64 | 2.46 |

With a computer powered by 4 CPUs, 3.40GHz, 16 GB RAM, the calculation time for Stage 1 is under 20 seconds. The calculation time for the rotation minimization is under 1 minute. The maximum calculation time for the deployment simulation is 6 minutes. Co optimization and parallelization are currently ongoing to enable applications in clinical conditions.

The results of Stage 1 (first and second steps) show that the SG position is globally well predicted in the artery, but that the radial deployment of stents is not sufficiently accurate, with errors exceeding the 3mm threshold. However, corrections made in Stage 2 (third and fourth step) yield a reasonable accuracy. More specifically, accurate predictions of fenestration positions after Stage 2 would avoid complex catheterization of secondary arteries, which is one of the major source of complications for practitioners.

However, the method has several limitations. First, it depends on the quality of the input data. Indeed, we assume that stents can be individualized using image processing. If this assumption is usually satisfied, overlapped stent can challenge it, for example, in the case of very pronounced angulation of the proximal neck of the aneurysm. In addition, the geometry of the aorta is assumed to be known, as it was updated using a non-rigid registration method previously. However, registration errors could occur and add to the other errors mentioned in the chapter. The non-rigid registration step also involves restriction for clinical applications. As we assume that we know the geometry of the aorta before the SG deployment simulation, the non-rigid registration must be performed each time the mobile C-arm position is changed, which implies to perform a DSA in order to visualize the aorta. The method should therefore only be used at key points in the procedure. Moreover, segmentation errors of the aorta volume are possible, but they are difficult to identify and probably have a minor influence on the results. A priori, the presence of thrombus is not a major problem for segmentation. Indeed, the aortic lumen is segmented during preoperative planning, not the aorta itself, and the SG is deployed in the lumen. Effects of thrombus on overall aortic stiffness are considered in the previous non-rigid registration step. But local variations in stiffness, due to thrombus, calcifications or surrounding tissues are not taken into account. Other limitations are related to the method itself. In order to save calculation time, the deployment of stents was simulated individually within rigid boxes. The rigid nature of the boxes is obviously a simplified mechanical behavior of the arterial wall but it did not induce significant errors when the simulations were compared with the post-operative CT scan. Finally, by simulating the individual deployment of stents, we do not consider the effect of the textile graft connecting them together. For example, a highly crimped stent can change the deployment diameter of the neighboring stent independently of the local geometry of the aorta. However, initial results show that these limitations do not hamper the accuracy more than what is compatible with clinical expectations. The influence of other assumptions should be investigated further with additional patient data.

This method would be particularly suited for complex cases of thoracoabdominal aortic aneurysms that require the use of fenestrated stent grafts. Indeed, when positioning fenestrations, the surgeon can potentially encounter difficulties. In most cases, the method is able to provide assistance to the surgeon. A few exceptions can challenge the method. The method is based on data extracted in the plane of the intraoperative image. Missing information along the projection axis are provided by simulations. If the main deployment axis of the stent graft was located along the projection axis, the method may have difficulty simulating the device. In practice, this situation seems very unlikely though.

# IV.   CONCLUSION

We have presented in this chapter a methodology for simulating SG deployment based on intraoperative images coupled with mechanical models. The algorithm consists of a series of successive steps with increasing precision. A first step simulates the positioning of the SG within the aorta using a simplified finite element model and the centerline of the artery. The second step is an axisymmetric geometric reconstruction of the stents. The third step minimizes the rotation of the stent around its main axis. Finally, the last step consists in deploying each stent individually within a deployment box extracted from the imaging and geometry of the aorta.

The results of combined Stage 1 and Stage 2 yield a reasonable accuracy. More specifically, accurate positioning of fenestration would facilitate catheterization of secondary arteries. But the method suffers from several limitations. First, it depends on the quality of the input data and the ability of image processing to distinguish stents. In addition, the geometry of the aorta is supposed to be known. Next, the wall of the artery is considered rigid when the stents are deployed. Lastly, textiles are not simulated. The influence of these limitations on the accuracy of simulations needs to be explored further using additional patient data.

Finally, the translation of our methodology seems promising and could be generalized to all operating rooms equipped with mobile C-arms to assist SG deployment using real-time simulations in the future.

# REFERENCES

[1]     A. Dua, S. Kuy, C. J. Lee, G. R. Upchurch, and S. S. Desai, "Epidemiology of aortic aneurysm repair in the United States from 2000 to 2010," *J. Vasc. Surg.*, vol. 59, no. 6, pp. 1512–1517, 2014.

[2]     M. S. Sajid, M. Desai, Z. Haider, D. M. Baker, and G. Hamilton, "Endovascular Aortic Aneurysm Repair (EVAR) Has Significantly Lower Perioperative Mortality in Comparison to Open Repair: A Systematic Review," *Asian J. Surg.*, vol. 31, no. 3, pp. 119–123, Jul. 2008.

[3]     R. M. Greenhalgh, L. C. Brown, J. T. Powell, S. G. Thompson, and D. Epstein, "Endovascular repair of aortic aneurysm in patients physically ineligible for open repair.," *N. Engl. J. Med.*, 2010.

[4]     S. Macdonald, R. Lee, R. Williams, and G. Stansby, "Towards safer carotid artery stenting: A scoring system for anatomic suitability," *Stroke*, 2009.

[5]     J.-N. Albertini *et al.*, "Aorfix Stent Graft for Abdominal Aortic Aneurysms Reduces the Risk of Proximal Type 1 Endoleak in Angulated Necks: Bench-Test Study," *Vascular*, vol. 13, no. 6, pp. 321–326, Nov. 2005.

[6]     J.-N. Albertini, J. . Macierewicz, S. . Yusuf, P. . Wenham, and B. . Hopkinson, "Pathophysiology of Proximal Perigraft Endoleak Following Endovascular Repair of Abdominal Aortic Aneurysms: a Study Using a Flow Model," *Eur. J. Vasc. Endovasc. Surg.*, vol. 22, no. 1, pp. 53–56, Jul. 2001.

[7]     A. Carroccio *et al.*, "Predicting iliac limb occlusions after bifurcated aortic stent grafting: Anatomic and device-related causes," *J. Vasc. Surg.*, vol. 36, no. 4, pp. 679–684, Oct. 2002.

[8]     F. Cochennec, J. P. Becquemin, P. Desgranges, E. Allaire, H. Kobeiter, and F. Roudot-Thoraval, "Limb Graft Occlusion Following EVAR: Clinical Pattern, Outcomes and Predictive Factors of Occurrence," *Eur. J. Vasc. Endovasc. Surg.*, vol. 34, no. 1, pp. 59–65, Jul. 2007.

[9]     S. Akpek, T. Brunner, G. Benndorf, and C. Strother, "Three-dimensional imaging and cone beam volume CT in C-arm angiography with flat panel detector," *Diagn Interv Radiol*, 2005.

[10]    T. Fagan, J. Kay, J. Carroll, and A. Neubauer, "3-D guidance of complex pulmonary artery stent placement using reconstructed rotational angiography with live overlay," *Catheter. Cardiovasc. Interv.*, vol. 79, no. 3, pp. 414–421, Feb. 2012.

[11]    S. A. M. Baert, E. B. Van de Kraats, T. Van Walsum, M. A. Viergever, and W. J. Niessen, "Three-Dimensional Guide-Wire Reconstruction from Biplane Image Sequences for Integrated Display in 3-D Vasculature," *IEEE Trans. Med. Imaging*, 2003.

[12]    M. Hoffmann *et al.*, "Semi-automatic catheter reconstruction from two views," in *Lecture Notes in Computer Science (including subseries Lecture Notes in Artificial Intelligence and Lecture Notes in Bioinformatics)*, 2012.

[13]    M. Hoffmann *et al.*, "Reconstruction method for curvilinear structures from

two views," in *Medical Imaging 2013: Image-Guided Procedures, Robotic Interventions, and Modeling*, 2013.

[14] M. Wagner, S. Schafer, C. Strother, and C. Mistretta, "4D interventional device reconstruction from biplane fluoroscopy," *Med. Phys.*, vol. 43, no. 3, pp. 1324–1334, Feb. 2016.

[15] P. Mortier *et al.*, "A Novel Simulation Strategy for Stent Insertion and Deployment in Curved Coronary Bifurcations: Comparison of Three Drug-Eluting Stents," *Ann. Biomed. Eng.*, vol. 38, no. 1, pp. 88–99, Jan. 2010.

[16] G. A. Holzapfel, M. Stadler, and T. C. Gasser, "Changes in the Mechanical Environment of Stenotic Arteries During Interaction With Stents: Computational Assessment of Parametric Stent Designs," *J. Biomech. Eng.*, vol. 127, no. 1, p. 166, Mar. 2005.

[17] F. Auricchio, M. Conti, M. De Beule, G. De Santis, and B. Verhegghe, "Carotid artery stenting simulation: From patient-specific images to finite element analysis," *Med. Eng. Phys.*, vol. 33, no. 3, pp. 281–289, Apr. 2011.

[18] S. De Bock *et al.*, "Filling the void: A coalescent numerical and experimental technique to determine aortic stent graft mechanics," *J. Biomech.*, vol. 46, no. 14, pp. 2477–2482, Sep. 2013.

[19] C. Kleinstreuer, Z. Li, and M. A. Farber, "Fluid-Structure Interaction Analyses of Stented Abdominal Aortic Aneurysms," *Annu. Rev. Biomed. Eng.*, vol. 9, no. 1, pp. 169–204, Aug. 2007.

[20] F. Auricchio, M. Conti, S. Marconi, A. Reali, J. L. Tolenaar, and S. Trimarchi, "Patient-specific aortic endografting simulation: From diagnosis to prediction," *Comput. Biol. Med.*, vol. 43, no. 4, pp. 386–394, May 2013.

[21] D. Perrin *et al.*, "Patient-specific numerical simulation of stent-graft deployment: Validation on three clinical cases," *J. Biomech.*, vol. 48, no. 10, pp. 1868–1875, 2015.

[22] D. Perrin *et al.*, "Deployment of stent grafts in curved aneurysmal arteries: toward a predictive numerical tool," *Int. j. numer. method. biomed. eng.*, vol. 31, no. 1, p. e02698, 2015.

[23] D. Perrin *et al.*, "Patient-specific simulation of stent-graft deployment within an abdominal aortic aneurysm," Jan. 2014.

[24] D. Perrin *et al.*, "Patient-specific simulation of endovascular repair surgery with tortuous aneurysms requiring flexible stent-grafts," *J. Mech. Behav. Biomed. Mater.*, vol. 63, pp. 86–99, Oct. 2016.

[25] S. Appanaboyina, F. Mut, R. Lohner, C. M. Putman, and J. R. Cebral, "Computational fluid dynamics of stented intracranial aneurysms using adaptive embedded unstructured grids," *Int. J. Numer. methods fluids*, 2008.

[26] L. Flórez Valencia, J. Montagnat, and M. Orkisz, "3D models for vascular lumen segmentation in MRA images and for artery-stenting simulation," *IRBM*, vol. 28, no. 2, pp. 65–71, Jun. 2007.

[27] L. Flórez-valencia *et al.*, "3D graphical models for vascular-stent pose simulation To cite this version : HAL Id : hal-00682926," vol. 13, no. 3, pp. 235–248, 2012.

[28] I. Larrabide, A. Radaelli, and A. Frangi, "Fast Virtual Stenting with Deformable Meshes: Application to Intracranial Aneurysms," Springer,

Berlin, Heidelberg, 2008, pp. 790–797.

[29]  I. Larrabide, M. Kim, L. Augsburger, M. C. Villa-Uriol, D. Rüfenacht, and A. F. Frangi, "Fast virtual deployment of self-expandable stents: Method and in vitro evaluation for intracranial aneurysmal stenting," *Med. Image Anal.*, vol. 16, no. 3, pp. 721–730, Apr. 2010.

[30]  E. Flore, I. Larrabide, L. Petrini, G. Pennati, and A. Frangi, "Stent deployment in aneurysmatic cerebral vessels: Assessment and quantification of the differences between Fast Virtual Stenting and Finite Element Analysis," Sep. 2009.

[31]  A. Bernardini *et al.*, "Deployment of self-expandable stents in aneurysmatic cerebral vessels: Comparison of different computational approaches for interventional planning," *Comput. Methods Biomech. Biomed. Engin.*, vol. 15, no. 3, pp. 303–311, 2012.

[32]  K. Spranger and Y. Ventikos, "Which spring is the best? Comparison of methods for virtual stenting," *IEEE Trans. Biomed. Eng.*, vol. 61, no. 7, pp. 1998–2010, 2014.

[33]  K. Spranger, C. Capelli, G. M. Bosi, S. Schievano, and Y. Ventikos, "Comparison and calibration of a real-time virtual stenting algorithm using Finite Element Analysis and Genetic Algorithms," *Comput. Methods Appl. Mech. Eng.*, vol. 293, pp. 462–480, Aug. 2015.

[34]  J. Zhong *et al.*, "Fast Virtual Stenting with Active Contour Models in Intracranical Aneurysm," *Sci. Rep.*, vol. 6, no. January, pp. 1–9, 2016.

[35]  S. Demirci *et al.*, "3D stent recovery from one x-ray projection," *Lect. Notes Comput. Sci.*, vol. 6891 LNCS, no. PART 1, pp. 178–185, 2011.

[36]  X.-Y. Zhou, J. Lin, C. Riga, G.-Z. Yang, and S.-L. Lee, "Real-Time 3-D Shape Instantiation From Single Fluoroscopy Projection for Fenestrated Stent Graft Deployment," *IEEE Robot. Autom. Lett.*, vol. 3, no. 2, pp. 1314–1321, Apr. 2018.

[37]  J.-Q. Zheng, X.-Y. Zhou, C. Riga, and G.-Z. Yang, "Real-time 3D Shape Instantiation for Partially-deployed Stent Segment from a Single 2D Fluoroscopic Image in Fenestrated Endovascular Aortic Repair," *IEEE Robot. Autom. Lett.*, pp. 1–1, 2019.

[38]  N. Demanget *et al.*, "Computational comparison of the bending behavior of aortic stent-grafts," *J. Mech. Behav. Biomed. Mater.*, vol. 5, no. 1, pp. 272–282, 2012.

[39]  A. F. Frangi, W. J. Niessen, K. L. Vincken, and M. A. Viergever, "Multiscale vessel enhancement filtering," Springer, Berlin, Heidelberg, 1998, pp. 130–137.

[40]  "The Vascular Modeling Toolkit," 2019. [Online]. Available: www.vmtk.org.

[41]  M. Groher, D. Zikic, and N. Navab, "Deformable 2D-3D Registration of Vascular Structures in a One View Scenario Supplementary Material Derivative of the Difference Measure," *IEEE Trans. Med. Imaging*, vol. 28, no. 6, pp. 847–860, 2009.

[42]  J.-Q. Zheng, X.-Y. Zhou, C. Riga, and G.-Z. Yang, "3D Path Planning from a Single 2D Fluoroscopic Image for Robot Assisted Fenestrated

Endovascular Aortic Repair," Sep. 2018.

[43]    A. Pionteck, B. Pierrat, S. Gorges, J.-N. Albertini, and S. Avril, "Finite-Element based Image Registration for Endovascular Aortic Aneurysm Repair," *Clin. Biomech.*, pp. 1–17.

[44]    D. T. Project Chrono, "Chrono: An Open Source Framework for the Physics-Based Simulation of Dynamic Systems." [Online]. Available: https://github.com/projectchrono/chrono.

[45]    A. Tasora *et al.*, "Chrono: An open source multi-physics dynamics engine," in *Lecture Notes in Computer Science (including subseries Lecture Notes in Artificial Intelligence and Lecture Notes in Bioinformatics)*, 2016, vol. 9611, pp. 19–49.

[46]    A. Tasora, "Euler-Bernoulli corotational beams in Chrono::Engine," pp. 1–12, 2016.

[47]    A. Tasora, M. Anitescu, S. Negrini, and D. Negrut, "A compliant visco-plastic particle contact model based on differential variational inequalities," *Int. J. Non. Linear. Mech.*, vol. 53, pp. 2–12, 2013.

[48]    P. A. Cundall and O. D. L. Strack, "A discrete numerical model for granular assemblies," *Géotechnique*, vol. 29, no. 1, pp. 47–65, Mar. 1979.